\documentclass[final,3p,times,twocolumn]{elsarticlee}
\biboptions{comma,sort&compress}
\usepackage{here}
\usepackage{graphicx, epsfig, amsmath, amssymb}
\def\nin{\noindent}

\newcommand{\be}{\begin{equation}}
\newcommand{\ee}{\end{equation}}
\newcommand{\ba}{\begin{eqnarray}}
\newcommand{\ea}{\end{eqnarray}}

\newcommand{\chpt}{$\chi$PT}
\newcommand{\cO}{{\cal O}}

\newcommand{\gevs}{~\mbox{GeV}^2}

\newcommand{\leff}{\ensuremath{\text{L}_{10}}^{\rm eff}}  
\newcommand{\ceff}{\ensuremath{\text{C}_{87}}^{\rm eff}}  

\begin{document}
\begin{frontmatter}
\title{Duality violation in QCD Sum Rules with the LR correlator}

\author[label1,label3]{Mart\'in Gonz\'alez-Alonso\corref{cor1}}
	\address[label1]{Departament de F\'{\i}sica Te\`orica and IFIC, Universitat de Val\`encia-CSIC, Apt. Correus 22085, E-46071 Val\`encia, Spain.}
	\address[label3]{Department of Physics, University of Wisconsin-Madison, Madison, WI 53706 USA.}
	\cortext[cor1]{Speaker}
	\ead{Martin.Gonzalez@ific.uv.es}
\author[label1]{Antonio Pich}
	\ead{Antonio.Pich@ific.uv.es}
\author[label2]{Joaquim Prades}
	\address[label2]{CAFPE and Departamento de F\'{\i}sica Te\'orica y del Cosmos,
	\\Universidad de Granada, Campus de Fuente Nueva, E-18002 Granada, Spain.}

\begin{abstract}
\noindent
We analyse the so-called violations of quark-hadron duality in Finite Energy Sum Rules (FESRs) with the LR correlator, through the study of the possible high-energy behavior of the LR spectral function, taking into account all known short-distance constraints and the experimental tau-decay data. In particular we show that the use of pinched weights (PWs) allows to determine with high accuracy the dimension six and eight contributions in the Operator-Product Expansion, $\cO_6= \left(-4.3^{+0.9}_{-0.7}\right)\cdot 10^{-3}~\mbox{GeV}^{6}$ and $\cO_8= \left(-7.2^{+4.2}_{-5.3}\right)\cdot 10^{-3}~\mbox{GeV}^{8}$ \cite{GonzalezAlonso:2010rn,GonzalezAlonso:2010xf}.
\end{abstract}

\end{frontmatter}

\section{Introduction}
\nin
QCD sum rules (QCDSRs) \cite{Shifman:1978bx,deRafael:1997ea} have been widely used during the last thirty years to study many important aspects of QCD. They constitute a very useful tool, that enables us with a powerful connection between QCD parameters and physical observables.

Here we will focus on QCD sum rules with the non-strange LR correlator $\Pi(q^2) \equiv \Pi_{ud,LR}^{(0+1)}(q^2)$ defined by
\ba
\label{eq:LRcorrelator2}
\!\!\!\!\!\!\!\!\Pi^{\mu\nu}_{ud,LR}(q)
\!\!\!\!&=&\!\!\!\!  i \!\!\int\!\! \mathrm{d}^4 x \; \mathrm{e}^{i q x} \, \langle 0 | T \left( L_{ud}^\mu(x) R_{ud}^\nu(0)^\dagger \right) | 0 \rangle\!\!\!\!\!\!\!\!\!\!\!\!\!\!\!\!\\
\!\!\!\!=&&\!\!\!\!\!\!\!\!\!\!\!\! (q^\mu q^\nu\!\!-\!\!g^{\mu\nu} q^2 )~\Pi^{(0+1)}_{ud,LR}(q^2) + g^{\mu\nu} q^2~\Pi^{(0)}_{ud,LR}(q^2) \, ,\nonumber
\ea
where $L_{ud}^\mu(x)\equiv \overline{u} \gamma^\mu (1-\gamma_5) d$ and $R_{ud}^\mu(x)\equiv \overline{u} \gamma^\mu (1+\gamma_5) d$.

In the deep euclidean region, the correlator can be calculated using the Operator-Product Expansion (OPE)
\ba
\label{eq:OPE}
\!\!\!\!\!\!\!\!\Pi^{\rm OPE}(s) \; =\; \sum_{k=3}\; \frac{C_{2k}(\nu)\,\langle O_{2k}\rangle(\nu)}{(-s)^{k}}\;
 \equiv\; \sum_{k=3}\; \frac{\mathcal{O}_{2k}}{(-s)^{k}}~,
\ea
where $\langle O_{2k}\rangle(\nu)$ are vacuum expectation values of operators with dimension $d=2k$ and $C_{2k}(\nu)$ their associated Wilson coefficients, that contain logarithmic dependences with $-s$. This correlator is an interesting object in the study of non-perturbative QCD because it vanishes identically to all orders in perturbation theory in the chiral limit and so its OPE contains only power-suppressed contributions from dimension $d=2k$ operators, starting at $d=6$, as already indicated in \eqref{eq:OPE}.

Therefore the OPE of the correlator is dominated by $\cO_6$ and $\cO_8$, two quantities that have been determined by several groups during the last decade with somewhat contradictory results. Most of these works are based on the use of QCDSRs with the LR correlator to extract the value of $\cO_{6,8}$ from hadronic tau data. Given that the data used by these groups is the same, the discrepancies have to come from the exact implementation of the QCDSR and the estimation of the associated errors.

\begin{figure}[tb]
\begin{center}
\includegraphics[width=7.5cm]{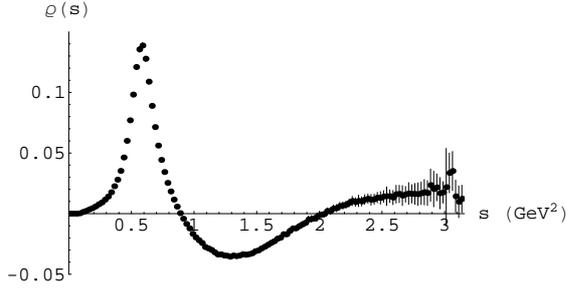}
\end{center}
\caption{Non-strange V-A spectral function $\rho(s)=\frac{1}{\pi}\mbox{Im}
\Pi^{(0+1)}_{ud,V-A}(s)$ measured from hadronic $\tau$ decays by ALEPH \cite{Schael:2005am}.}
\label{fig:VmenosA}
\end{figure}

A QCD Sum Rule takes advantage of the analytic properties of the correlator to relate its imaginary part in the positive real axis (where hadrons lie) with its value in the rest of the complex plane, where the OPE allows us to calculate it in terms of quarks and gluons.
We can write a general QCDSR for the LR correlator as
\ba
\label{eq:FESRwithDV}
&&\!\!\!\!\!\!\!\!\!\!\!\!\!\!\!\!\!\!\!\!\!\!\!\int^{s_0}_{s_{\rm th}} \!\!\!\mathrm{d}s\; w(s) \,\rho(s) + \frac{1}{2 \pi i} \oint_{|s|=s_0} \!\!\!\! \mathrm{d}s\; w(s) \,\Pi^{\rm{OPE}}(s) +  \mathrm{DV}[w,s_0] \nonumber\\
&& = 2 f_\pi^2\, w(m_\pi^2) ~ + ~ \underset{s=0}{\text{Res}} \left[ w(s) \, \Pi(s)\right] ,
\ea
where $\rho(s)\equiv\frac{1}{\pi}\,\mathrm{Im}\Pi(s)$ is the LR non-strange spectral function that has been measured in $\tau$ decays \cite{Schael:2005am,Ackerstaff:1998yj,Barate:1998uf} (see Fig.~\ref{fig:VmenosA}) and w(s) is an arbitrary weight function that is analytic in the whole complex plane except in the origin (where it can have poles). The r.h.s. contains the pion-pole contribution and the residue at the origin for negative-power weight functions, $1/s^n$, which is calculable with Chiral Perturbation Theory ($\chi$PT).
The quark-hadron duality violation (DV) comes from the breakdown of the OPE near the positive real axis  \cite{Poggio:1975af}, and using analyticity it can be written in the form \cite{Shifman:2000jv,Chibisov:1996wf,Cata:2005zj,GON07}
\ba
\label{eq:DV2}
{\rm DV} [w(s),s_0]  ~=~ \int^{\infty}_{s_0} \mathrm{d}s~w(s)~\rho(s)~,
\ea
that shows the DV as the part of the integral of the spectral function that is not included in the sum rule.

\section{Extracting $L_{10}^{\mathrm{eff}}$, $C_{87}^{\mathrm{eff}}$, $\mathcal{O}_{6,8}$: FESRs and PWs}
\nin
We will analyse the DV effects in different QCDSR extractions of 
$L_{10}^{\mathrm{eff}}\equiv -\overline{\Pi}(0)/8$, $C_{87}^{\mathrm{eff}}\equiv \overline{\Pi}'(0)/16$, $\mathcal{O}_{6}$ and $\mathcal{O}_{8}$,
where $\overline{\Pi}(s)\equiv \Pi (s)-2 f_\pi^2/(s-m_\pi^2)$. 
The first two quantities, that can be expressed in terms of low-energy \chpt~constants \cite{GonzalezAlonso:2008rf}, are very well known, with good agreement between the different phenomenological and theoretical determinations and the DV contribution is expected to be small. On an opposite situation we have $\cO_{6,8}$, that are defined in Eq.~\eqref{eq:OPE}, especially for $\cO_8$ where the different works do not even agree in its sign. The DV is much larger for these last two quantities.

The simplest sum rules that can be used to extract their value are the FESRs obtained with the weights $w(s)=s^n$, with $n=-2,-1,2,3$:\footnote{We neglect here the logarithmic corrections to the Wilson coefficients in the OPE. The associated error is expected to be smaller than other errors in the analysis, as was found e.g. in Refs.~\cite{Cirigliano:2003kc,Ciulli:2003sc}.}
\ba
\label{eq:C87}
\!\!\!\!\!\!\!\!\!\int^{s_0}_{s_{\rm th}} \mathrm{d}s\; \frac{1}{s^2} \, \rho(s)\;
&=&\; 16 \, C_{87}^{\rm eff}\, - \, \rm{DV}[1/s^2,s_0]\, , \\
\label{eq:L10}
\!\!\!\!\!\!\!\!\!\int^{s_0}_{s_{\rm th}} \mathrm{d}s\; \frac{1}{s} \, \rho(s)\;
&=&\; -8 L_{10}^{\rm eff}\, - \, \rm{DV}[1/s, s_0]\, , \\
\label{eq:M2}
\!\!\!\!\!\!\!\!\!\int^{s_0}_{s_{\rm th}} \mathrm{d}s\; s^2 \, \rho(s)\;
&=&\; 2 f_\pi^2 m_\pi^4\, +\,\cO_6 \, - \, \rm{DV}[s^2,s_0]\, , \\
\label{eq:M3}
\!\!\!\!\!\!\!\!\!\int^{s_0}_{s_{\rm th}} \mathrm{d}s\; s^3 \, \rho(s)\;
&=&\; 2 f_\pi^2 m_\pi^6\, -\, \cO_8\, - \, \rm{DV}[s^3,s_0]\, .
\ea

A more refined strategy makes use of the pinched-weights (PWs), polynomial weights that vanish at $s=s_0$ and are supposed to minimize the DV \cite{LeDiberder:1992fr,Cirigliano:2002jy,Dominguez:2003dr,Cirigliano:2003kc,Bordes:2005wv}. However Eq.~\eqref{eq:DV2} shows that things are more subtle \cite{Cata:2005zj,GON07,GonzalezAlonso:2010rn} and that it depends on the particular weight used and on how fast the spectral function goes to zero. In our particular case and avoiding the introduction of unknown condensates of higher dimension, we have the following PW sum rules:
\ba
\label{eq:C87pw}
&& \!\!\!\!\!\!\!\!\!\!\!\!\!\!\!\!\!\!\!\!\! \int^{s_z}_{s_{\rm th}} \!\mathrm{d}s\; \frac{\rho(s)}{s^2}\left( 1-\frac{s}{s_z}\right)^2 \left(  1+\frac{2s}{s_z}\right)
= ~16~C_{87}^{\rm eff} - 6 \frac{f_\pi^2}{s_z^2}~,\\
\label{eq:L10pw}
&& \!\!\!\!\!\!\!\!\!\!\!\!\!\!\!\!\!\!\!\!\! \int^{s_z}_{s_{\rm th}} \!\mathrm{d}s~\frac{\rho(s)}{s}\left( 1-\frac{s}{s_z}\right)^2
= -8 L_{10}^{\rm eff} - 4 \frac{f_\pi^2}{s_z}~,\\
\label{eq:M2pw}
&& \!\!\!\!\!\!\!\!\!\!\!\!\!\!\!\!\!\!\!\!\! \int^{s_z}_{s_{\rm th}} \!\mathrm{d}s~  \rho(s)\left(s - s_z \right)^2
~=\, 2 f_\pi^2 s_z^2 +\cO_6 \,,\\ 
\label{eq:M3pw}
&& \!\!\!\!\!\!\!\!\!\!\!\!\!\!\!\!\!\!\!\!\! \int^{s_z}_{s_{\rm th}} \!\mathrm{d}s~\rho(s) \left(s - s_z \right)^2 \left(s +2 s_z \right)
= 4f_\pi^2 s_z^3 - \cO_8\, , 
\ea
where the DV contribution and the negligible terms proportional to the pion mass have not been explicitly written for the sake of brevity.

\section{Estimating the quark-hadron duality violation}
\nin
In Ref. \cite{GonzalezAlonso:2010rn} we have studied the DV from the perspective given by Eq.~\eqref{eq:DV2}, using the parametrization
\ba
\label{eq:model}
\rho(s\ge s_z) &=& \kappa~ e^{-\gamma s} \sin(\beta (s-s_z))~,
\ea
for the spectral function beyond $s_z \sim 2.1\gevs$ and finding the region in the 4-dimensional parameter space that is compatible with the most recent experimental data \cite{Schael:2005am} and the following theoretical constraints: first and second Weinberg Sum Rules \cite{Weinberg:1967kj} (WSRs) and the sum rule of Das et al. \cite{Das:1967it} that gives the electromagnetic mass difference of pions ($\pi$SR).
The parametrization (\ref{eq:model}) emerges naturally in a resonance-based model \cite{Blok:1997hs,Shifman:1998rb,Shifman:2000jv} and has been used recently to study violations of quark-hadron duality \cite{Cata:2005zj,GON07,Cata:2008ye,Cata:2008ru}, although without imposing the previously explained theoretical constraints in the numerical analysis.

\begin{figure}[ht]
\vfill
\centerline{
\begin{minipage}[t]{.3\linewidth}\centering
\centerline{\includegraphics[width=3.8cm]{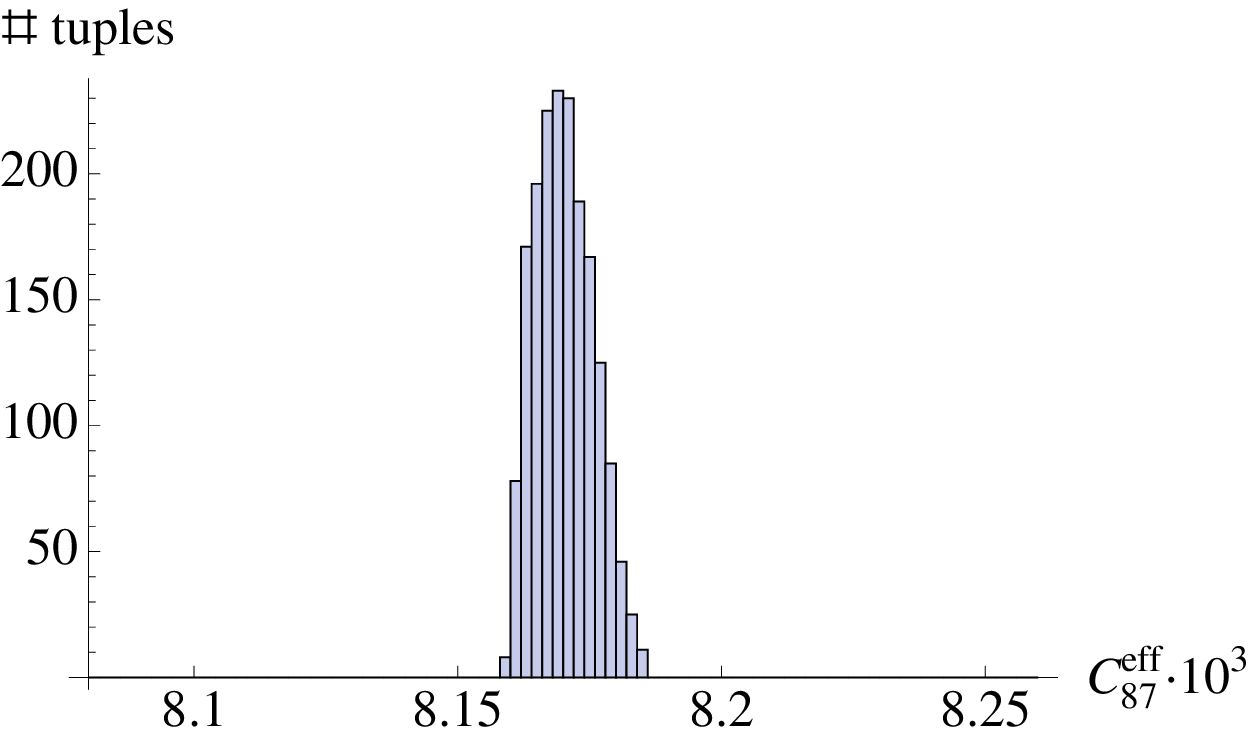}}
\end{minipage}
\hspace{1.4cm}
\begin{minipage}[t]{.3\linewidth}\centering
\centerline{\includegraphics[width=3.8cm]{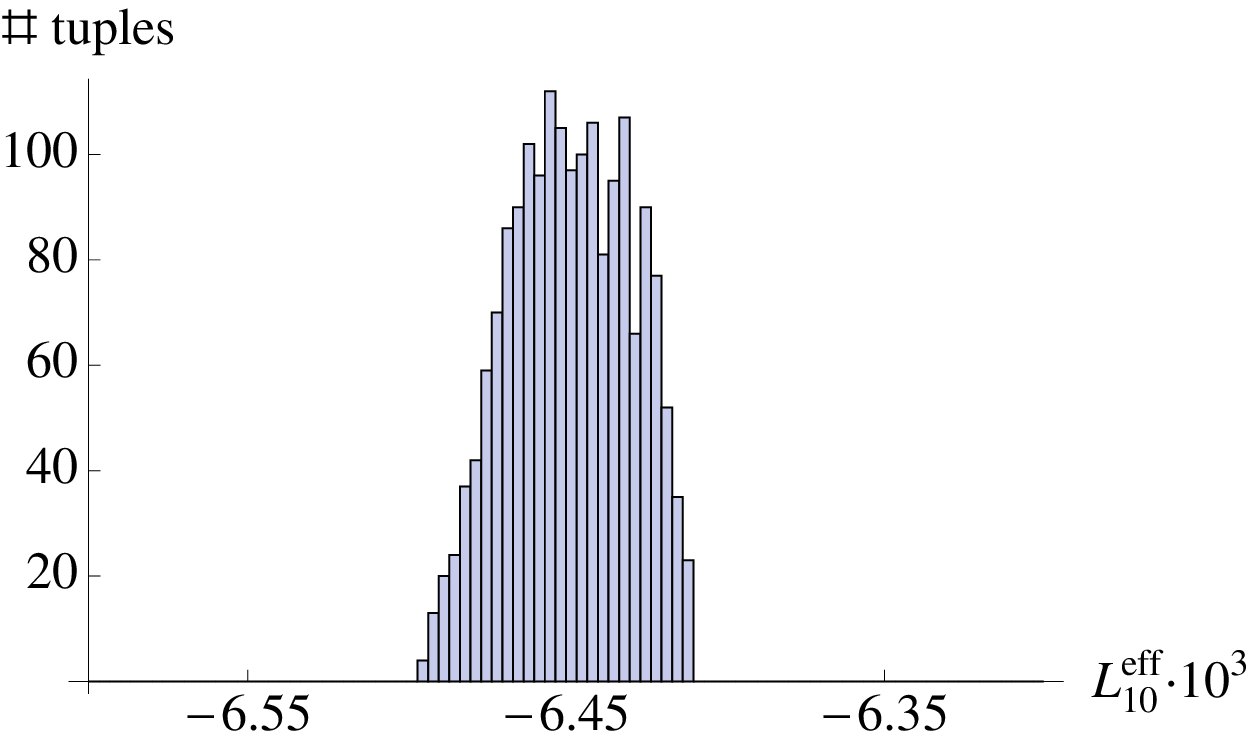}}
\end{minipage}
}
\vspace{0.7cm}
\centerline{
\begin{minipage}[t]{.3\linewidth}\centering
\centerline{\includegraphics[width=3.8cm]{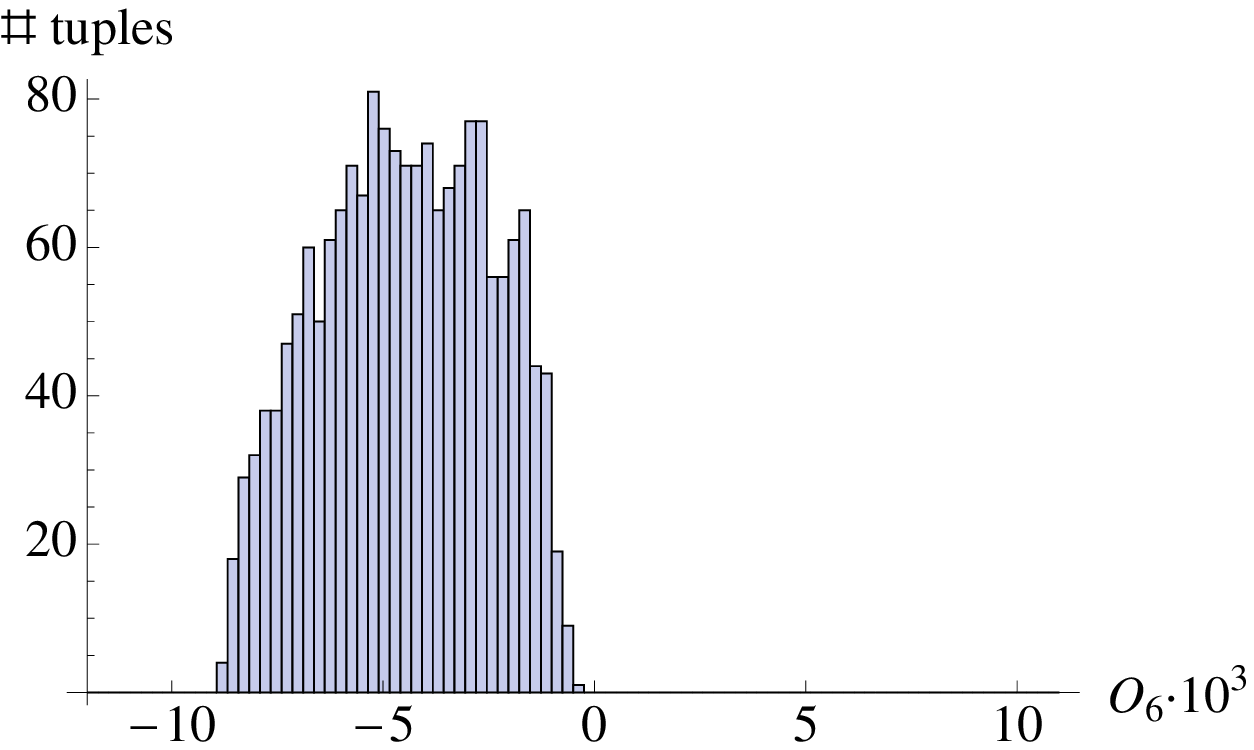}}
\end{minipage}
\hspace{1.4cm}
\begin{minipage}[t]{.3\linewidth}\centering
\centerline{\includegraphics[width=3.8cm]{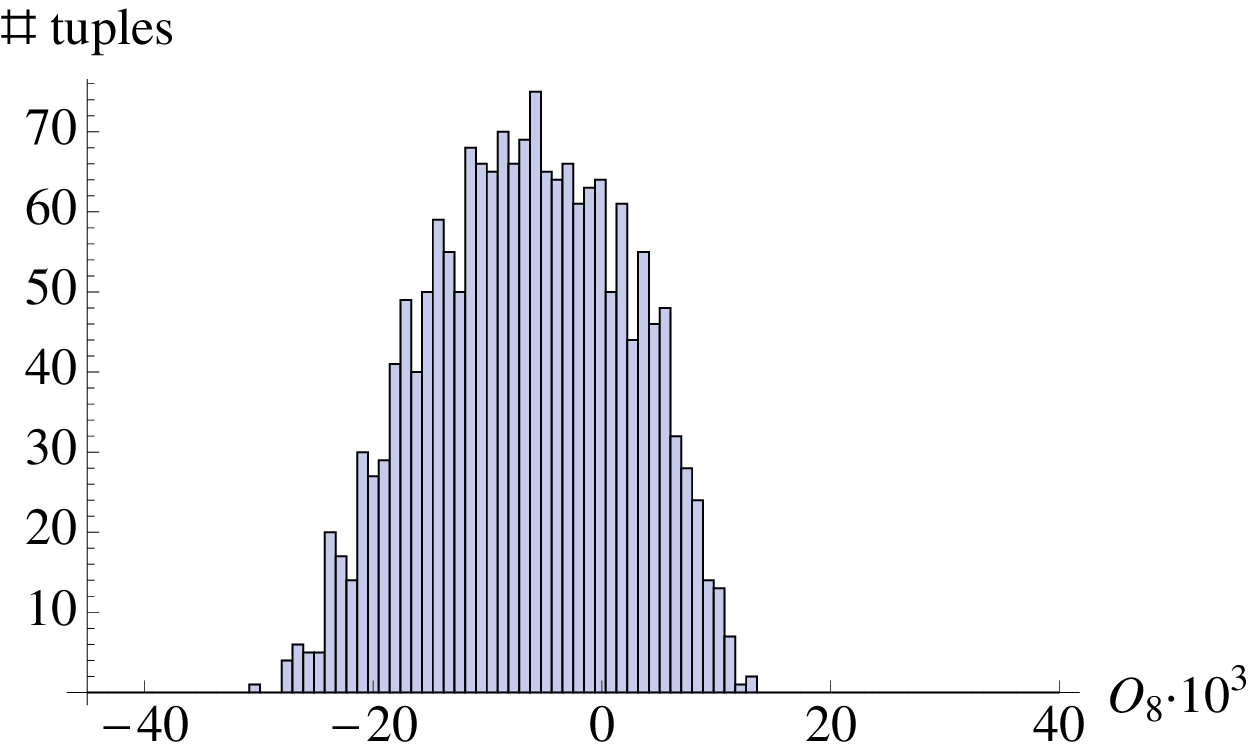}}
\end{minipage}
}
\vfill
\caption{Statistical distribution of values of $C_{87}^{\mathrm{eff}}$ (upper-left), $L_{10}^{\mathrm{eff}}$ (upper-right), $\cO_6$ (lower-left) and $\cO_8$ (lower-right)  for the accepted spectral functions, using the sum rules \eqref{eq:C87} - \eqref{eq:M3}. The parameters are expressed in GeV to the corresponding power.}
\label{fig:observables}
\end{figure}

Performing a numerical scanning over the parameter space $(\kappa,\gamma,\beta,s_z)$, 
we have generated a large number of "acceptable" spectral
functions, satisfying all conditions, and have used them to extract the wanted hadronic
parameters.
Carrying out the integrals in Eqs.~(\ref{eq:C87}--\ref{eq:M3}) with $s_0 \to \infty$,
one obtains the results summarized in Fig.~\ref{fig:observables}, which shows the statistical distribution of the calculated parameters \cite{GonzalezAlonso:2010rn}.

From these distributions, one gets the final numbers (for the $68\%$ probability region):
\ba
\label{eq:C87result1}
C_{87}^{\mathrm{eff}} &=&
\left(8.17\,\pm0.12\right)\cdot 10^{-3}~\mbox{GeV}^{-2}~ , \\ \label{eq:L10result}
L_{10}^{\mathrm{eff}}&=&
\left(-6.46\, {}^{+\, 0.08}_{-\, 0.07}\right)\cdot 10^{-3}\, , \\ \label{eq:O6result}
\cO_6&=&
\left(-5.4\, {}^{+\, 3.6}_{-\, 1.6}\right)\cdot 10^{-3}~\mbox{GeV}^6\, , \\ \label{eq:O8result1}
\cO_8&=&
\left(-8.9\, {}^{+\, 12.6}_{-\, 7.4}\right)\cdot 10^{-3}~\mbox{GeV}^8\, ,
\ea
where the error includes both the DV and the experimental contributions.

\begin{figure}[ht]
\vfill
\centerline{
\begin{minipage}[t]{.3\linewidth}\centering
\centerline{\includegraphics[width=3.8cm]{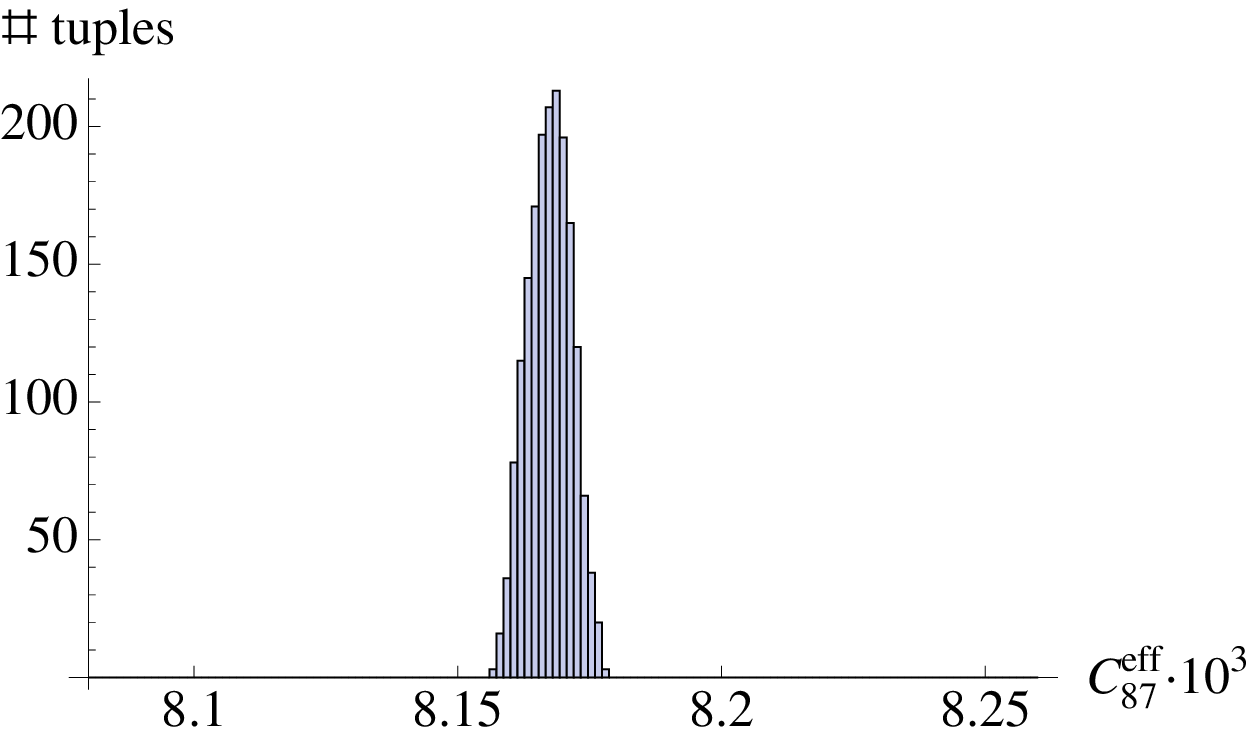}}
\end{minipage}
\hspace{1.4cm}
\begin{minipage}[t]{.3\linewidth}\centering
\centerline{\includegraphics[width=3.8cm]{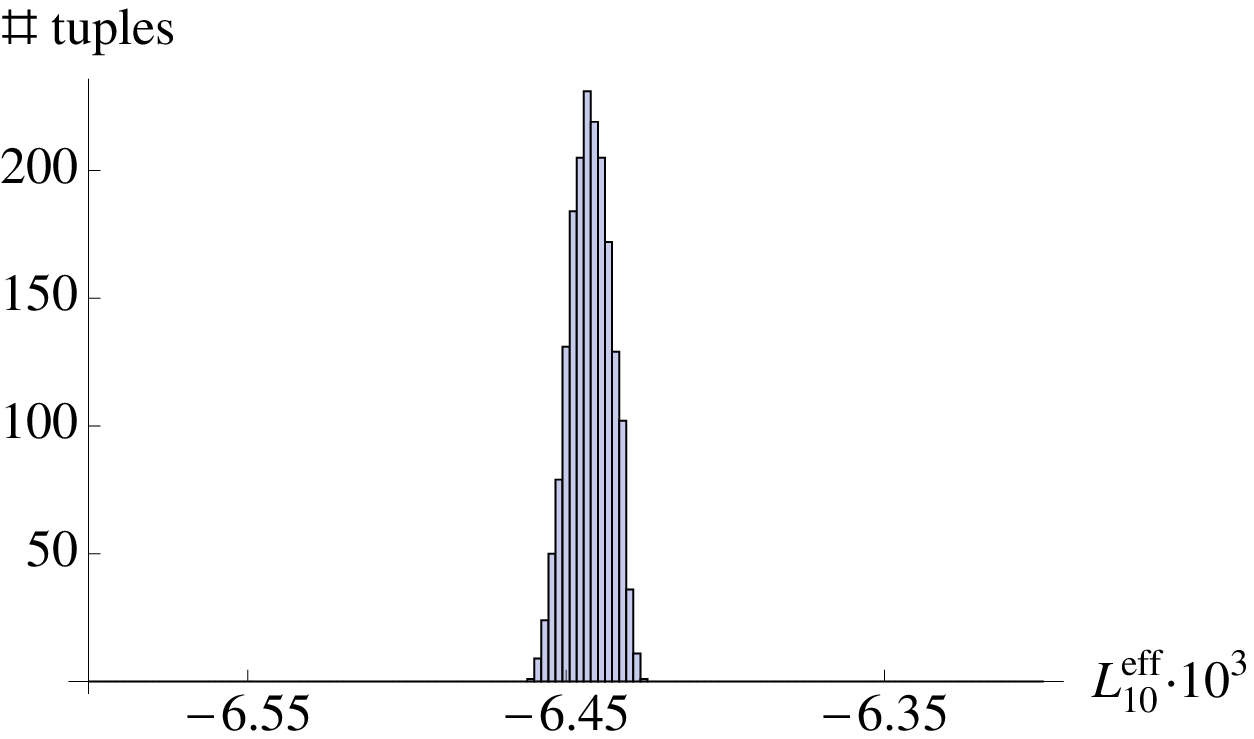}}
\end{minipage}
}
\vspace{0.7cm}
\centerline{
\begin{minipage}[t]{.3\linewidth}\centering
\centerline{\includegraphics[width=3.8cm]{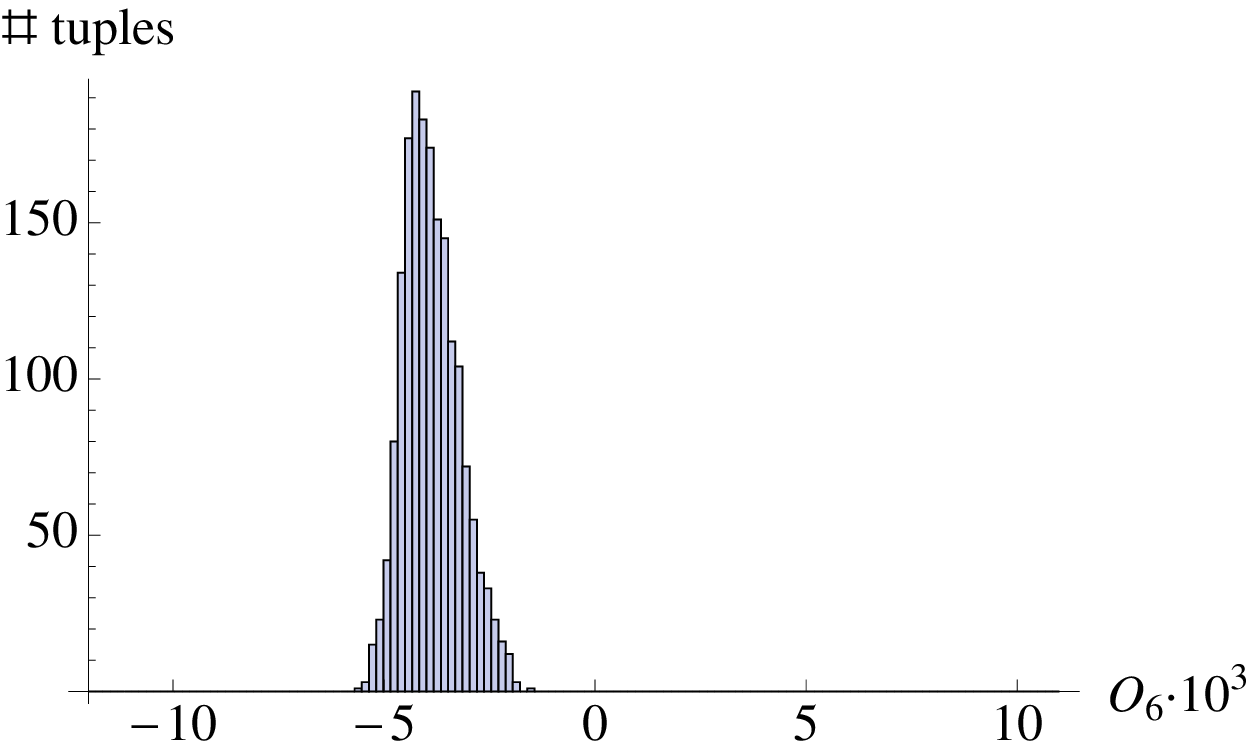}}
\end{minipage}
\hspace{1.4cm}
\begin{minipage}[t]{.3\linewidth}\centering
\centerline{\includegraphics[width=3.8cm]{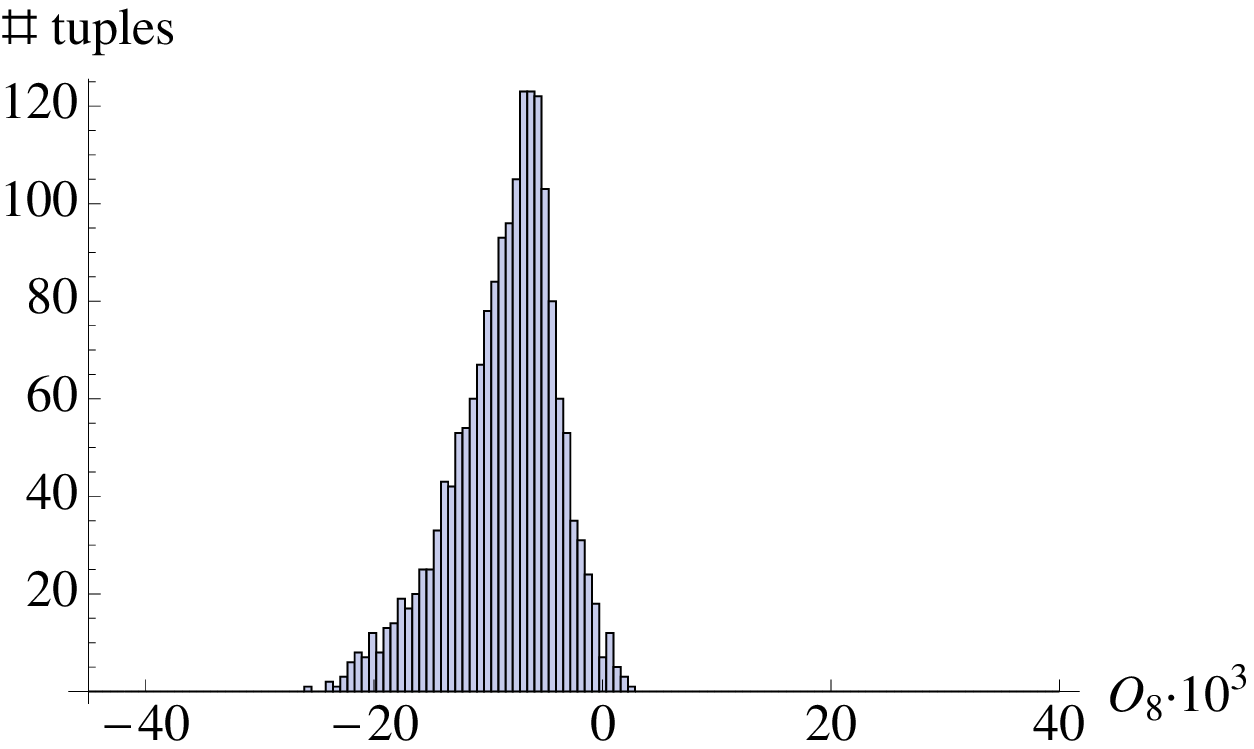}}
\end{minipage}
}
\vfill
\caption{Statistical distribution of values of $C_{87}^{\mathrm{eff}}$ (upper-left), $L_{10}^{\mathrm{eff}}$ (upper-right), $\cO_6$ (lower-left) and $\cO_8$ (lower-right) for the accepted spectral functions, using the pinched-weight sum rules \eqref{eq:C87pw} - \eqref{eq:M3pw}. The parameters are expressed in GeV to the corresponding power.}
\label{fig:observablesPW2}
\end{figure}

In  Ref.~\cite{GonzalezAlonso:2010xf} we have applied the same procedure to study the PW sum rules, Eqs.~(\ref{eq:C87pw}--\ref{eq:M3pw}), obtaining the results shown in Fig. \ref{fig:observablesPW2}. We can see that the histograms are much more peaked around their central values than those obtained in Ref.~\cite{GonzalezAlonso:2010rn} with standard weights.
%

The corresponding numerical values are
\ba
\label{eq:C87resultPW68}
\!\!\!\!\!\ceff \cdot\! 10^3~\mbox{GeV}^2 \!\!\!\!&=&\!\!\!\! 8.17\pm 0.12 = 8.17 \pm 0.24~,\\
\label{eq:L10resultPW68}
\!\!\!\!\!\leff \cdot\! 10^3 \!\!\!\!&=&\!\!\!\! -6.44 \pm 0.05 =-6.4 \pm 0.1~,\\
\label{eq:O6resultPW68}
\!\!\!\!\!\cO_6  \cdot\! 10^3~\mbox{GeV}^{-6} \!\!\!\!&=&\!\!\!\! -4.3\, {}^{+0.9}_{-0.7} = -4.3\, {}^{+2.1}_{-1.5}~,\\
\label{eq:O8resultPW68}
\!\!\!\!\!\cO_8  \cdot\! 10^3 ~\mbox{GeV}^{-8} \!\!\!\!&=&\!\!\!\! -7.2\, {}^{+4.2}_{-5.3}=-7.2\, {}^{+8.6}_{-12.7}~,
\ea
where the first and second results correspond to the $68\%$ and $95\%$ probability regions respectively. The error includes both the DV and the experimental contributions.


\section{Conclusions and comparisons}
Using a physically motivated model for the DV and through a procedure detailed in Ref.~\cite{GonzalezAlonso:2010rn}, we have studied the possible high-energy behavior of the LR spectral function, once the most recent experimental data \cite{Schael:2005am} and the known theoretical constraints have been imposed.

\begin{figure}[ht]
\vfill
\centerline{
\begin{minipage}[t]{.3\linewidth}\centering
\centerline{\includegraphics[width=7.35cm]{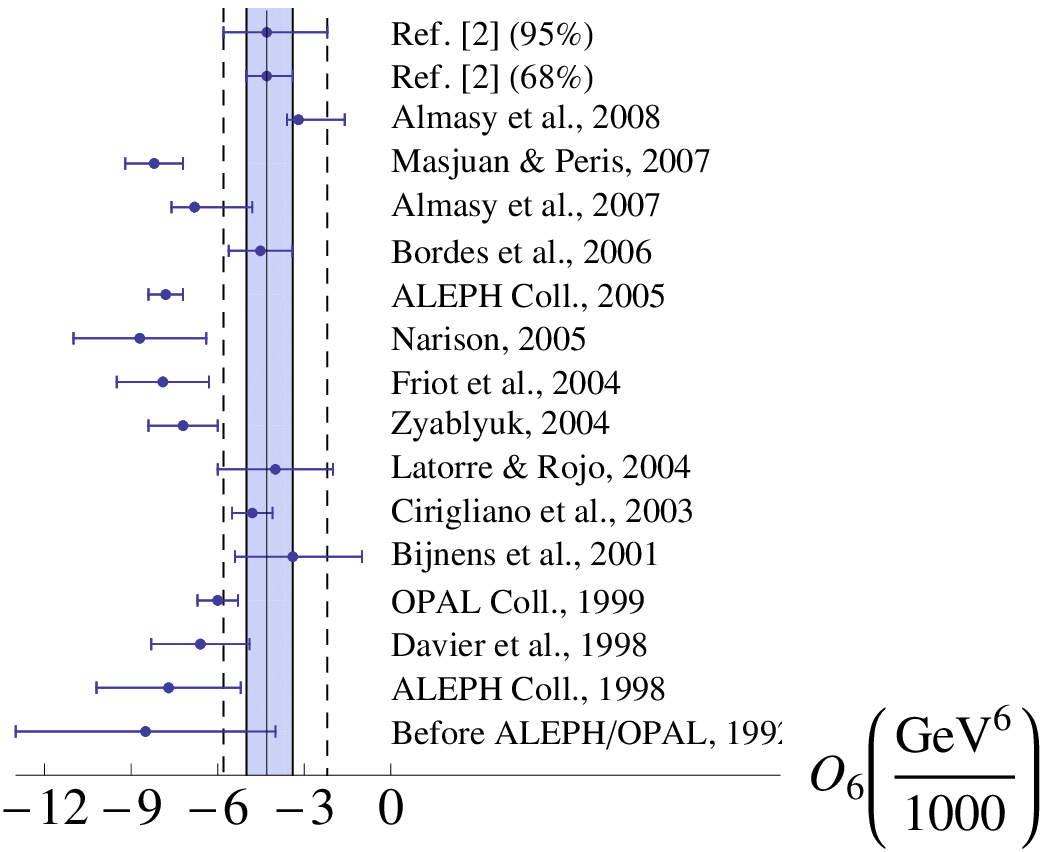}}
\end{minipage}
}
\vspace{0.75cm}
\centerline{
\begin{minipage}[t]{.3\linewidth}\centering
\centerline{\includegraphics[width=7.35cm]{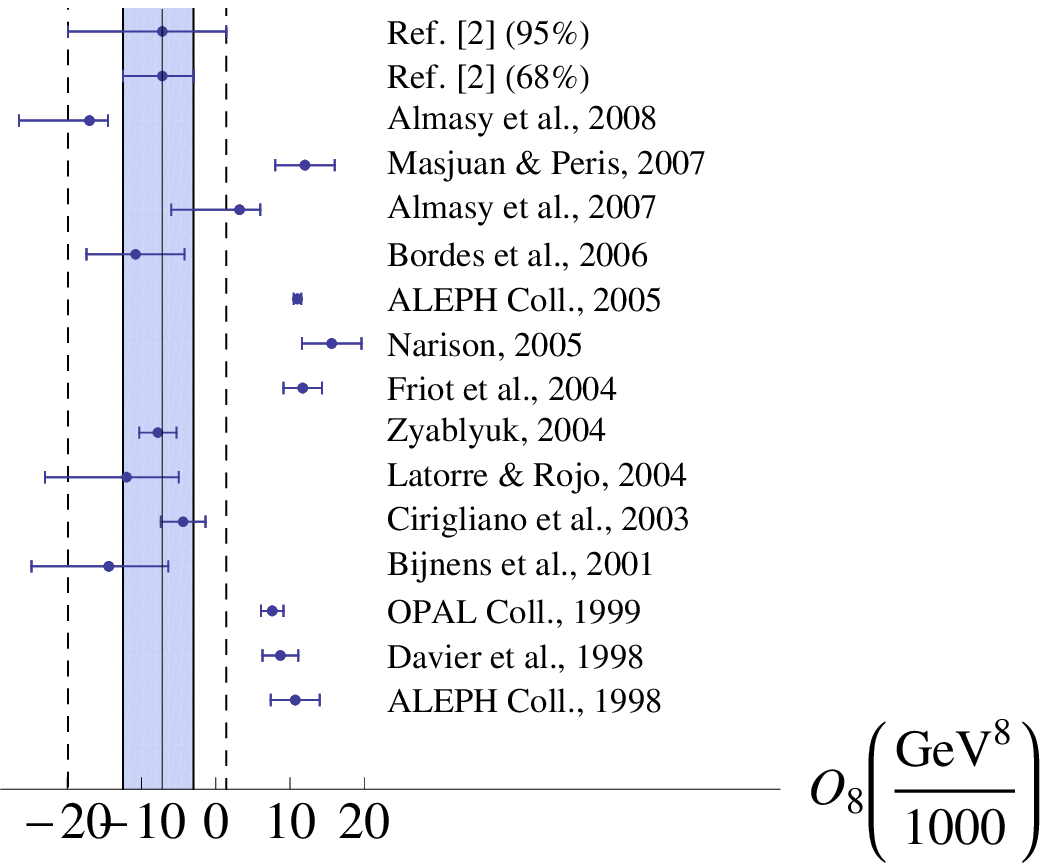}}
\end{minipage}
}
\vfill
\caption[]{Comparison of our results for $\cO_6$ (upper plot) and $\cO_8$ (lower plot) with previous determinations \cite{Schael:2005am,Ackerstaff:1998yj,Barate:1998uf,Davier:1998dz,Bijnens:2001ps,Cirigliano:2002jy,Cirigliano:2003kc,Ioffe:2000ns,Zyablyuk:2004iu,Rojo:2004iq,Narison:2004vz,Dominguez:2003dr,Bordes:2005wv,Peris:2000tw,Friot:2004ba,Almasy:2006mu,Masjuan:2007ay,Almasy:2008xu}.
}
\label{fig:comparisonPW2}
\end{figure}

In this way we have analyzed the error of different standard and pinched-weight FESRs and we have extracted the value of several hadronic parameters.  Our results for the low-energy constants $\leff$ and $\ceff$ are in excellent agreement with the precise determination of Ref.~\cite{GonzalezAlonso:2008rf}.
%
We have determined the condensates $\cO_6$ and $\cO_8$ using the PW sum rules \eqref{eq:M2pw} and \eqref{eq:M3pw}, checking that the PW succeeds in minimizing the errors and concluding that the most recent experimental data provided by ALEPH, together with the theoretical constraints (WSRs and $\pi$SR), fix with accuracy the value of $\cO_6$ and determine the sign of $\cO_8$. Our results are compared in Fig.~\ref{fig:comparisonPW2} with previous determinations\footnote{In Fig.~\ref{fig:comparisonPW2} we have taken into account that in the $\cO_8$ determination of Refs. \cite{Ioffe:2000ns,Zyablyuk:2004iu} there was a sign error, as was pointed out in Ref.~\cite{Ioffe:2005ym}.}.

It can be seen that our results agree with those of Refs.~\cite{Cirigliano:2002jy,Cirigliano:2003kc,Dominguez:2003dr,Bordes:2005wv}, based also on pinched weights, Ref.~\cite{Bijnens:2001ps}, based on the second duality point, and with Ref.~\cite{Rojo:2004iq} that follows a technique similar to ours. Our analysis indicates that the DV error associated to the use of the first duality point is very large and was grossly underestimated in Refs.~\cite{Zyablyuk:2004iu,Narison:2004vz}. In Refs.~\cite{Friot:2004ba,Masjuan:2007ay,Cata:2009fd} the numerical values obtained at this first duality point are supported through theoretical analyses based on the so-called ``minimal hadronic ansatz'' or Pad\`e approximants, but our results show that the first duality point is very unstable when we change from the WSRs to the $\cO_{6,8}$ sum rules, indicating that the systematic error of these approaches is non-negligible. Essentially the same can be said about Refs.~\cite{Barate:1998uf,Davier:1998dz} where the last available point $s_0=m_\tau^2$ was used.

Summarizing, our results agree within two standard deviations with previous estimates of $\cO_6$ and they indicate that $\cO_8$ is also negative.

\section*{Acknowledgements}
\nin
This work is dedicated to the memory of our collaborator and friend Ximo Prades, who sadly passed away. This work has been partly supported by the EU network FLAVIAnet [MRTN-CT-2006-035482], by MICINN, Spain  [FPA2007-60323, FPA2006-05294 and CSD2007-00042 --CPAN--], by Generalitat Valenciana [Prometeo/2008/069] and by Junta de Andaluc\'{\i}a [P07-FQM 03048 and P08-FQM 101].
%
%
\providecommand{\href}[2]{#2}\begingroup\raggedright
\endgroup

\end{document}